\documentclass{article}
\usepackage{graphicx}
\usepackage{subcaption}
\usepackage{arxiv}
\usepackage{diagbox}
\usepackage[utf8]{inputenc} 
\usepackage[T1]{fontenc}    
\usepackage{hyperref}       
\usepackage{url}    
\usepackage{hyperref}
\usepackage{booktabs}       
\usepackage{amsfonts}       
\usepackage{nicefrac}       
\usepackage{microtype}      
\usepackage{lipsum}
\usepackage{xcolor}
\usepackage{algorithm}
\usepackage{indentfirst}
\usepackage{multirow}
\usepackage[noend]{algpseudocode}
\usepackage{tikz,siunitx,mwe}
\usepackage{diagbox}
\title{Deriving information from missing data: implications for mood prediction}

\author{
  Yue Wu \\
  Mathematical Institute\\
  University of Oxford\\
  Oxford,  OX2 6GG, UK; \\
  Alan Turing Institute\\
  London, NW1 2DB, UK \\
  \texttt{yue.wu@maths.ox.ac.uk} \\
      \And 
   Terry J. Lyons \\
 Mathematical Institute\\
  University of Oxford\\
  Oxford, OX2 6GG, UK; \\
  Alan Turing Institute\\
  London, NW1 2DB, UK \\
  \texttt{terry.lyons@maths.ox.ac.uk}  \\
     \And
Kate E.A. Saunders\\
Department
of Psychiatry
\\University of Oxford; \\
  Oxford Health NHS Foundation Trust\\
  Warneford Hospital\\
  Oxford OX3 7JX, UK \\
  \texttt{kate.saunders@psych.ox.ac.uk} \\
}

\begin{document}
\maketitle

\begin{abstract}
The availability of mobile technologies has enabled the efficient collection prospective longitudinal, ecologically valid self-reported mood data from psychiatric patients. These data streams have potential for improving the efficiency and accuracy of psychiatric diagnosis as well predicting future mood states enabling earlier intervention. However, missing responses are common in such datasets and there is little consensus as to how this should be dealt with in practice. A signature-based method was used to capture different elements of self-reported mood alongside missing data to both classify diagnostic group and predict future mood in patients with bipolar disorder, borderline personality disorder and healthy controls. The missing-response-incorporated signature-based method achieves roughly 66\% correct diagnosis, with f1 scores for three different clinic groups 59\% (bipolar disorder), 75\% (healthy control) and 61\% (borderline personality disorder) respectively. This was significantly more efficient than the naive model which excluded missing data. Accuracies of predicting subsequent mood states and scores were also improved by inclusion of missing responses.  The signature method provided an effective approach to the analysis of prospectively collected mood data where missing data was common and should be considered as an approach in other similar datasets.
\end{abstract}


\section{Introduction}
\label{sec:Introduction}

The rapid emergence of mobile technologies has transformed the way in which mental health data can be collected. Until recently clinicians were wholly reliant on anamnestic approaches and thus hampered by the inaccuracy of retrospective recall regarding psychiatric symptoms. Mobile technologies have enabled the efficient capture of self-reported symptoms in an ecologically valid and prospective manner.  A number of different approaches to the analysis of longitudinal mood data have been employed \cite{faurholt2015daily,bopp2010longitudinal,arribas2018signature}. However missing data is ubiquitous and poses a significant methodological challenge. Mood data may be missing unrelated to mood state or in fact be a consequence of current mood state. Standard approaches such as imputation may inadvertently lead to the loss of important information \cite{faurholt2019guideline}. 

Signatures from rough path theory \cite{lyons2002rough} are an effective way of analyzing these types of data as they capture the order in which events occur. The approach incorporates missing data into the system in order to capture the underlying patterns and the evolving interactions between missing data and responses, and to predict the future from the past of an evolving system \cite{levin2013past}. So far, the signature method has significantly contributed to automated recognition of Chinese handwriting \cite{gaham2013sparse,xie2017learning}, formulation of appropriate stochastic partial differential equations to model randomly evolving interfaces \cite{hairer2013kpz, hairer2014regularity}, skeleton-based human action recognition \cite{yang2017skeleton}, diagnosis of Alzheimer’s disease \cite{moore2019using} and speech emotion recognition \cite{wang2019speech}. In a previous analysis we demonstrated that a signature-feature model could be successfully applied to 6-dimensional self-reported mood data \cite{arribas2018signature}, however missing data was not used in this analysis.

In this study, we used a missing-response-incorporated signature-feature-based machine learning model to re-analyse weekly mood data collected from the AMoSS study \cite{tsanas2016truecolor} which used self-reported mood data and wearables to distinguish between individuals with bipolar disorder (BD), borderline personality disorder (BPD) and healthy controls (HC). We sought to test whether this new analytic approach was superior to standard approaches to mood quantification in its ability to distinguish these diagnostic groups and predict future mood states/scores.

\section{Methods}
\label{sec:Methods}
\subsection{Data}
 \label{sec:data}
Patients with BD or BPD and healthy volunteers reported their mood using Altman Self-Rating Mania scale (ASRM) \cite{altman1997asrm} and the Quick Inventory of Depressive Symptoms (QIDS-SR16) \cite{rush2003qids} were collected. ASRM is a short, 5-item self-assessment questionnaire assessing the presence and severity of manic or hypomanic symptoms. QIDS-SR16 contains 16 items covering the 9 DSM-IV symptom criterion domains \cite{apa2013dsm} with the total score ranging from 0 to 27.

The data were collected as part of the AMoSS study \cite{tsanas2016truecolor}. ASRM and QIDS data were collected from 142 individuals as part of the AMoSS study and the participants completed standardised questionnaires on a weekly basis using the True Colors mood monitoring system \cite{goodday2008monitor} after receiving a text or email prompt. Two of the 142 participants either withdrew consent or had no clinical diagnosis and were therefore excluded from analysis. We further excluded 14 participants who failed to provide at least 20 weeks data as part of the analysis is based on information in data of at least 20 weeks. Of the remaining 126 participants, 49 were diagnosed as bipolar disorder and 32 were borderline personality disorder. All identical duplicate values were checked and removed, and only the first response of a week was kept if multiple responds happened within that week. Most of the time, assessments were completed one after the other, of which the sum scores can then be paired up. We discarded the excessive observed data which could not be paired. Using this data, 2-dimensional concatenated data of paired observations from ASRM and QIDS was obtained for each participant, where the score is ‘-1’ for a missing response.

\begin{center}
    \begin{table}[!htbp]
\caption { Demographic characteristics of the three clinical groups (the appropriate distributions are summarised in the form of the median $+/-$ in the interquartile range).} \label{tab:demographic} 
\centering
\bgroup
\def\arraystretch{1.2}
\vspace{0.3cm}
\begin{tabular}{cccccc}
\toprule[1.5pt]
\midrule
Group & Recruited & for analysis   & Weeks in study & Ages & Gender (males)\\
\midrule
BD & 54 & 49   &  51$\pm$3 & 38$\pm$19 & 16\\
HC & 52 &   45 &  51$\pm$2 &  37$\pm$18& 13\\
BPD & 34 & 32  & 51$\pm$2 &  34$\pm$12 & 3\\

\midrule
\bottomrule[1.25pt]
\end{tabular}
\egroup
\end{table}

\end{center}
\subsection{Features extraction}
    
\subsubsection{Encoding missing data}
Among all the 126 valid participants in our study, 90\% missed a response on a least one occasion during their task-active weeks. Signature features allow missing responses to be included in the analysis without the need for imputation. The missing events are translated into a new counting process \cite{little2019missing}. An example is illustrated in Figure 1 for the procedure. In the general case, if one work on data with N many time points, the accumulative missing counts can be generated for each of the $N$ time points by calculating the sum of missing observations up to that particular time point; meanwhile each missing observation, i.e., input "-1" in our case, is replaced by the valid value that happened in the nearest past (referred as the \emph{feed forward} method). This does not imply that the missing responses are assumed to take the same value as their nearest valid responses. By doing this, the increments in both observation and missing counts can be preserved and captured, which are indeed the most critical features in the signature method together with their functionals \cite{lyons2014rough}.

After transforming missing responses, one then normalises (and accumulate like in \cite{arribas2018signature}) the data to make it scale-free in order to apply signature transformation.

\begin{figure}[htbp]
    \centering
        \includegraphics[clip, trim=0.5cm 25.5cm 6.5cm 2.5cm, width=0.80\textwidth]{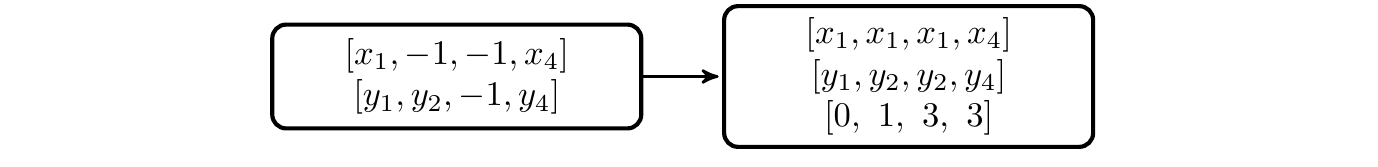}
    \caption{An example for integrating missing responses: the left block contains 2 dimensional data of 4 consecutive observations, where -1 represents one missing observation; in the right block the feed forward method is applied for filling in all missing places with valid values that happened in the corresponding nearest past, while an additional dimension is added to count missing events cumulatively at each time points.}
    \label{fig:example1}
\end{figure}

\subsubsection{Signature features}
In recent year, the signatures of the continuous paths generated from longitudinal data is considered as an efficient feature set for learning purpose because of its nature to capture the nonlinear effect of the evolving systems \cite{lyons2014rough}. Consider $\mathbb{R}^d$-valued time-dependent, piecewise-differentiable paths of finite length. Such a path $X$ mapping from time domain $[a,b]$ to $\mathbb{R}^d$ is denoted as $X:[a,b] \to \mathbb{R}^d$. For short we will use $X_t$ for $X(t), t\in [a,b]$. Each coordinate path of $X$ is a real-valued path and denoted as $X^i, i\in [d]$ with $[d]:=\{1,\ldots,d\}$. The \emph{signature} of a path $X:[a,b]\to \mathbb{R}^d$, denoted by $S(X)_{a,b}$, is the infinite collection of all iterated integrals of $X$. That is, 
\begin{equation}\label{eqn:signature}
    S(X)_{a,b}:=(1,S(X)_{a,b}^1, \ldots,S(X)_{a,b}^d, S(X)_{a,b}^{1,1}, S(X)_{a,b}^{1,2},\ldots),
\end{equation}
where, the $0$th term is 1 by convention, and the superscripts of the terms after the $0$th term run along the set of all multi-index
$\{(i_1,\ldots,i_k)|k\geq 1, i_1,\ldots,i_k \in [d]\}$ with the coordinate iterated integral being
 \begin{equation}\label{eqn:iterated_integral}
S(X)_{a,b}^{i_1,\ldots,i_{k}}:=\int_{a<t_{k}<b}\cdots \int_{a<t_{1}<t_2} \text{d}X_{t_1}^{i_1}\ldots \text{d}X_{t_k}^{i_k}.
\end{equation}
The finite collection of all terms $S(X)_{a,b}^{i_1,\ldots,i_k}$ with the multi-index of fixed length $k$ is termed as the \emph{kth level of the signature}. The truncated signature up to the $p$th level is denoted by $\lfloor S(X)_{a,b} \rfloor_p$. In machine learning context, truncated signature features are always obtained by truncating the original signature to some finite level.

For discrete data stream $\mathbf{x}=\big(\mathbf{x}_1,\ldots,\mathbf{x}_n\big)$, where $\mathbf{x}$ contains $n$ observations, and the $i$th observation $\mathbf{x}_i$, $i\in [n]$, is assumed to be a $d$-dimensional column vector at the $i$th time point, one needs to convert it to a continuous $\mathbb{R}^d$-valued path via piecewise linear interpolation or other transforms in order to compute signature. The availability of Python packages \emph{iisignature} \cite{reizenstein2018iisignature} and \emph{esig} allows easy calculation of signature, where the linear interpolation is implemented automatically by the packages.

For our purpose, we extracted the consecutive paired observations for each participant, incorporated the missing data, and then calculated the corresponding signature features via Python package iisignature, where the signature features were truncated to level 2. To distinguish from standard signature features, our features were named the missing-response-incorporated signature features (MRSF).

\subsection{Signature-based classification} \label{sec: model_class}
In order to investigate the role of ASRM and QIDS scores in differentiating between healthy controls and different patient groups, a missing-response-incorporated signature-based classification model was developed to classify the diagnostic group a participant belonged to. For each of 126 participants, a stream of 20 consecutive paired observations, no matter missing or not, was randomly drawn from the 2-dimensional concatenated data for this task. Then the collection of the 20 consecutive paired data were randomly split into a training set (70\%) and a testing set (30\%). The proposed model was based on a random forest classifier and was trained on the MRSF of the training set. 

For comparison, the random forest classifier was also trained on features extracted through a clinic-used metric based on the average score in each category over the valid scores in 20 consecutive observations (the naive classification model). The performance of the missing-response-incorporated signature-based classification model (MRSCM, level 2) and the baseline model (the naive classification model), for classifying the diagnostic groups were tested on bootstrapping and measured in terms of accuracy. Meanwhile the confusion matrices of both methods were generated to allow more detailed analysis, from which f1 scores for different clinic groups were computed. To assess the separation ability of different methods, we created the receiver operating characteristic curves (ROC) at various threshold settings and computed areas under curve (auc).

\subsubsection{Spectrum analysis}\label{sec:spectrum1}
To further test the performance of the MRSCM (level 2), we investigated the likelihood of each of the three groups being categorised into the correct clinic group. We trained the model on the 20 consecutive streamed data of all but one participant and tested it with the data of this participant. The probability vector of each participant being classified into each group was calculated and then projected onto the equilateral triangle, with each vertex representing one of the three clinic groups. For example, if the inferred probabilities of one participant being classified as BD, HC and BPD are 0.1, 0.5 and 0.4 respectively, then the corresponding probability vector is $[0.1,0.5,0.4]$. This vector is indeed on a 3-dimensional triangle surface $[p,q,1-p-q]$, with non-negative $p,q$ and $p+q\leq 1$. This triangle is the equilateral triangle that all the inferred 3-dimensional probability vectors will be sitting on.

\subsection{Signature-based predictions}\label{sec: model_pred}
We then sought to predict the mania and depression states/scores of each participant in the next week on the last 10 paired observations. Empirically, using the last 10 consecutive data was more effective than using 20, perhaps the future mood state/score is dependent on the most recent states. 
\subsubsection{State prediction}
At this stage, in order to simplify the prediction problem, we reduced the outcome to one of three mutually exclusive responses. We treated ASRM and QIDS similarly but separately. That is, for ASRM, no answer (future response is missing), normal (score of ASRM is no bigger than 5) or manic (score of ASRM is bigger than 5); and for QIDS, no answer (future response is missing), normal (score of QIDS is no bigger than 10) or depressed (score of QIDS is bigger than 10). 

The target of the prediction is not quantified, so we approached the problem as a classification problem rather than as a regression. We chose to learn the relation between the pattern of last 10 consecutive observations and the future state through classifiers. For this task, the MRSF truncated at level 2 were extracted from data and a random forest classifier-based predictive model was trained for each clinical group separately. For comparison, the random forest classifier-based predictive model was also trained on features extracted through computing the mean of the last 10 consecutive observations (naive predictive model). The prediction of ASRM/QIDS states of each participant based on diagnostic group using the missing-response-incorporated signature-based predictive model (stateMRSPM, level 2) and the naive predictive model were tested with bootstrapping and measured in terms of accuracy. 

\subsection{Spectrum analysis}
To assist the understanding towards the distribution of the responses to ASRM/QIDS, the bar histograms of the proportions of selected features, namely, no answer/normal/manic for ASRM and no answer/normal/depressed for QIDS, for each clinic group were plotted for an overview.

We also visualised the true proportion vectors as well as the prediction results of each participant using triangle spectrum plots. This is a natural adoption as we also have three different states: no answer, normal and manic/depressed. 

The true proportion vector reflects the ground truth. It consists of three elements, that is, the proportion of this participant giving 'no answer', 'normal' and 'manic' (resp. 'depressed')  for ASRM (resp. QIDS) during his/her entire study. In order to demonstrate group-dependent characteristics, true proportion vectors for the same questionnaire of patients from the same clinical group were visualised in the same 3-dimensional equilateral triangle surface.

Regarding the predicted result, for each participant of one of the three groups (BD/HC/BPD), ASRM/QIDS states of 5 weeks rather than 1 week were predicted using stateMRSPM (level 2) trained from random 10 consecutive observations of the rest participants in the same group. To avoid overfitting, we excluded the participants that generated at most 5 buckets of 10 consecutive weeks data. Then the predicted proportion of his/her giving each of the three different responses for future ASRM/QIDS could be directly calculated from the predicted ASRM/QIDS states of the five weeks. 

\subsection{Score prediction}
We also sought to predict the next reported ASRM/QIDS score made by a participant based on features extracted from their previous 10 weekly observations, no matter missing or not. To ensure comparability of our results, we predicted raw ASRM/QIDS scores. For our own comparison, we also made more coarse-grained predictions according to the categories for QIDS: none (1-5), mild (6-10), moderate (11-15), severe (16-20) and very severe (21-27) \cite{rush2003qids}, and for ASRM: none (0-5),  mild (6-9), moderate (10-13), severe (14-17) and very severe (17-20).

For this task, the MRSF truncated at level 2 were extracted from each sequential data and a random forest regressor-based predictive model was trained for each clinical group separately. The future score prediction of each participant using missing-response-incorporated signature-based predictive model (scoreMRSPM, level 2) was tested with bootstrapping and measured in terms of accuracy and mean absolute error (MAE).
\subsection*{Summary}
We used the publicly available Python iisignature package (version 0.23) to calculate signatures of streams of data, Python numpy package (version 1.19.0) for data manipulations and processing, Python scikit-learn package (version 0.23.1) for machine learning tasks and matplotlib for plotting and graphics (version 3.2.1).

The study was approved by the NRES Committee East of England—Norfolk (13/EE/0288) and the Research and Development department of Oxford Health NHS Foun- dation Trust.

A summary of tasks and models can be found in Table \ref{tab:tasksummary} and Table \ref{tab:modelsummary}.

\section{Results}
\label{sec:results}

\subsection{Classification of the diagnostic group}
\label{sec:group}
MRSCM (level 2) categorised 66.3\% of participants into the correct with a standard deviation 0.06 while the naive model only classified 54.9\% of participants correctly with a higher variability 0.08. The accuracy from MRSCM improved with transformation of missing responses, indicating that missing responses bring additional information and therefore enhance the performance of the model. 

\begin{figure}[!htbp]
  \centering
\begin{subfigure}[b]{0.47\linewidth}
\includegraphics[width=\linewidth]{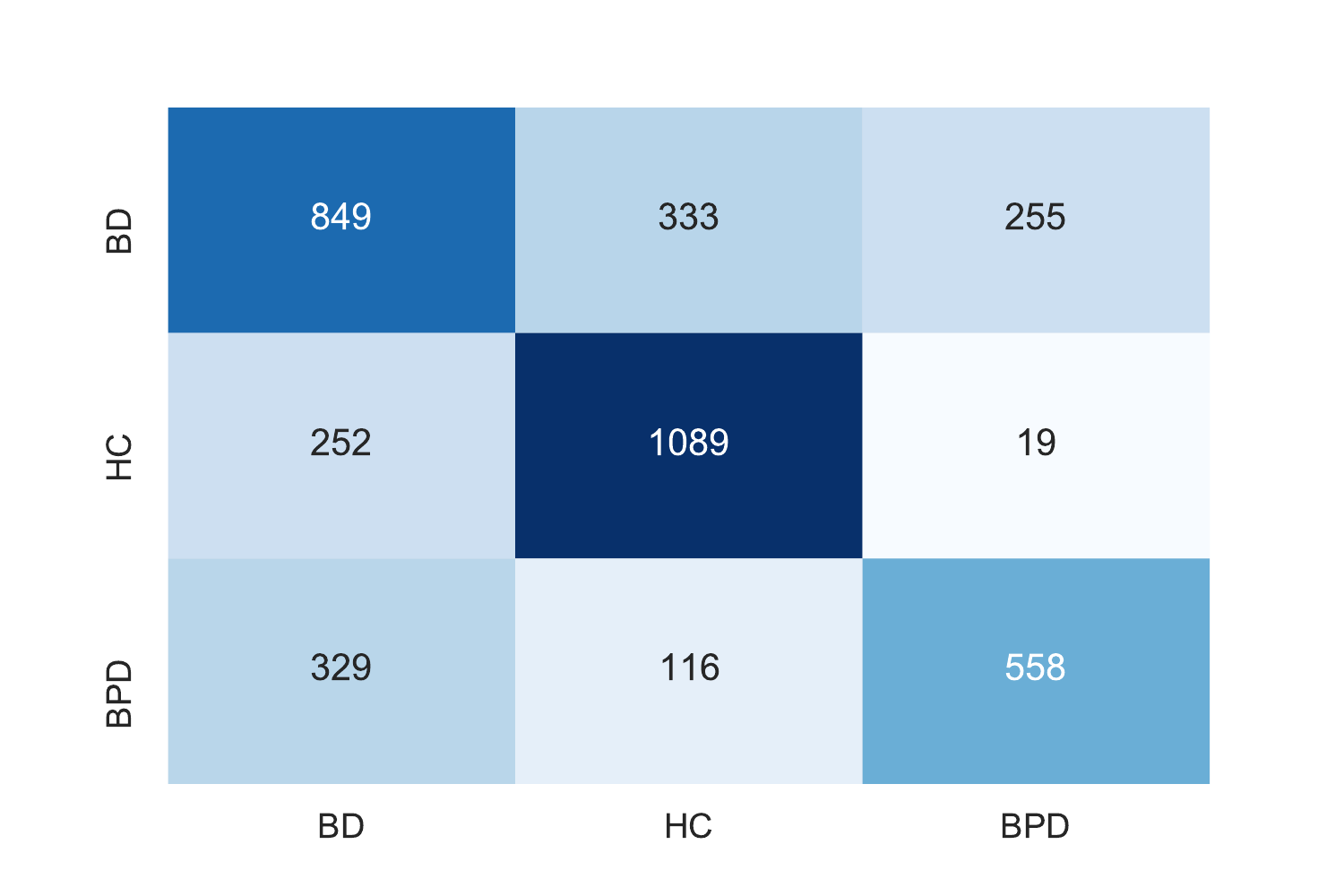}
\caption{MRSCM (level 2).}
\end{subfigure}
\begin{subfigure}[b]{0.47\linewidth}
\includegraphics[width=\linewidth]{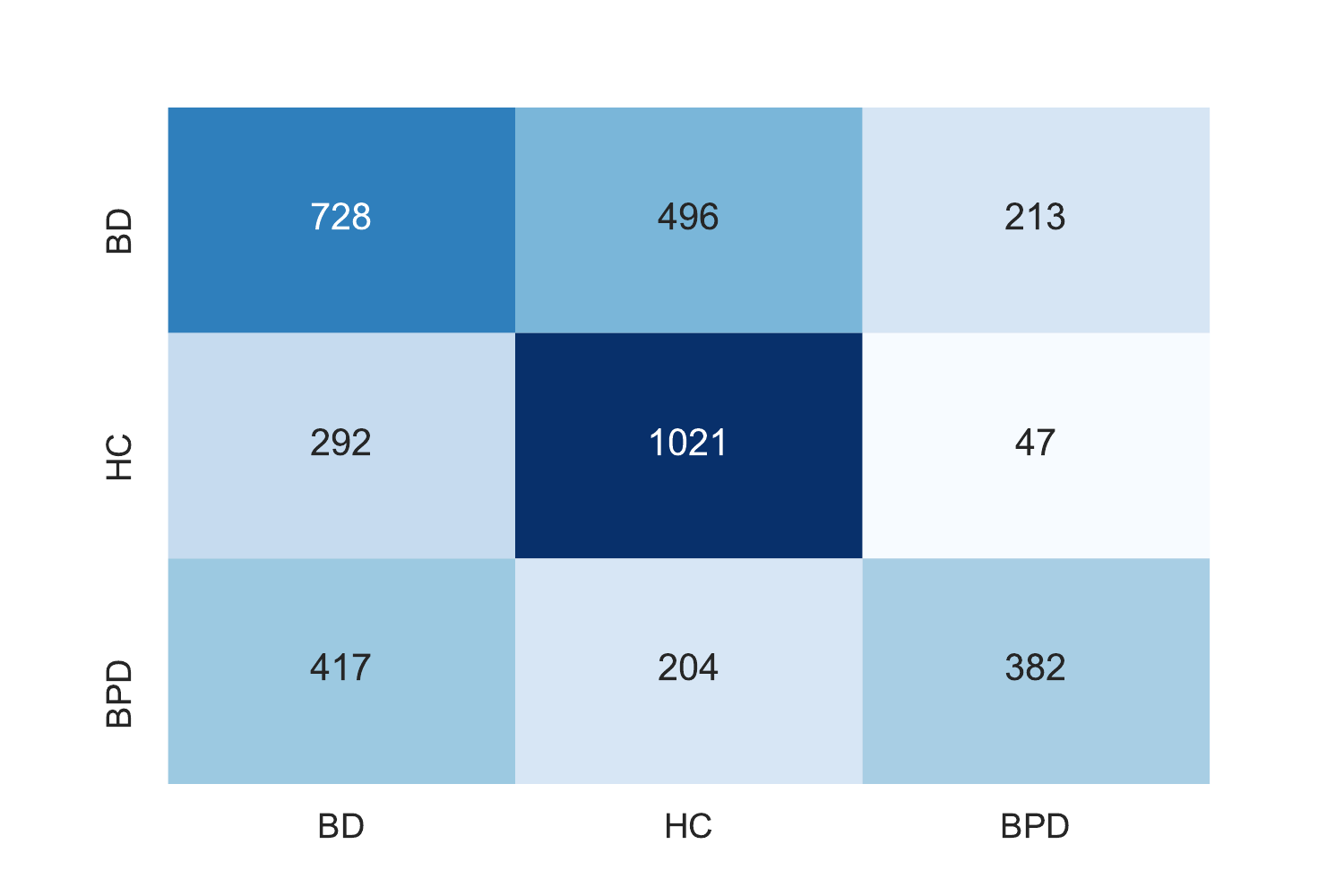}
\caption{Naive classification model.}
\end{subfigure}
\caption{Confusion matrices of the missing-response-incorporated signature-based classification model (MRSCM, level 2) and the naive classification model.}
\label{fig:confusion}
\end{figure}

Figure \ref{fig:confusion} gives the confusion matrices for both methods, which illustrates the detailed correct and false classification for each group and allows for computing f1 scores in Table \ref{tab:f1}. Table \ref{tab:f1} shows that the MRSCM (level 2) has higher f1 scores in all three classes. It achieved its lowest f1 score for classifying bipolar group. However, encoding the missing information into the model, the ability of classifying BPD was significantly enhanced from 40\%+ (the naive model) to 60\%+ (MSRCM, level 2). 

\begin{figure}[!htbp]
  \centering
\begin{subfigure}[b]{0.47\linewidth}
\includegraphics[width=\linewidth]{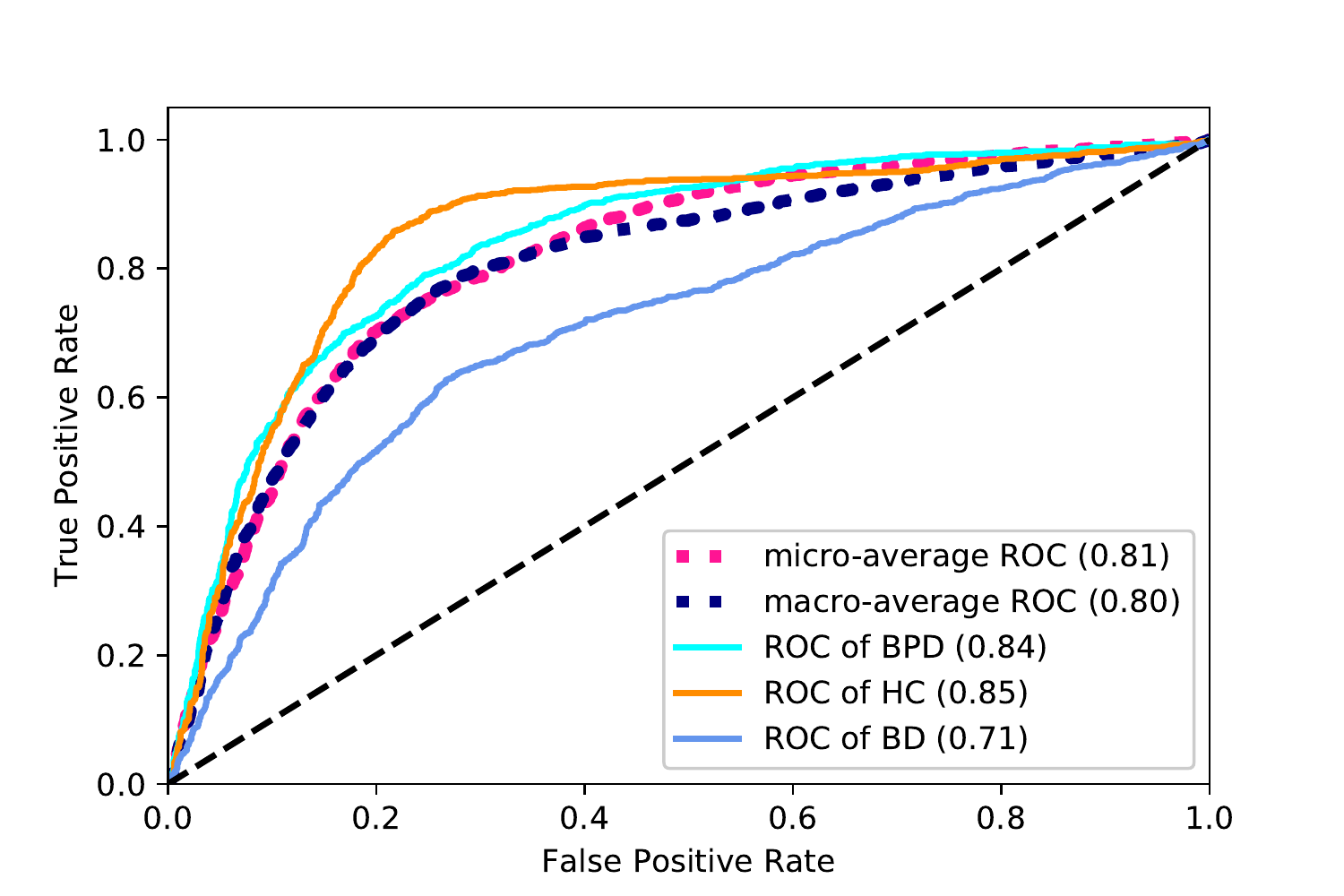}
\caption{MRSCM (level 2).}
\end{subfigure}
\begin{subfigure}[b]{0.47\linewidth}
\includegraphics[width=\linewidth]{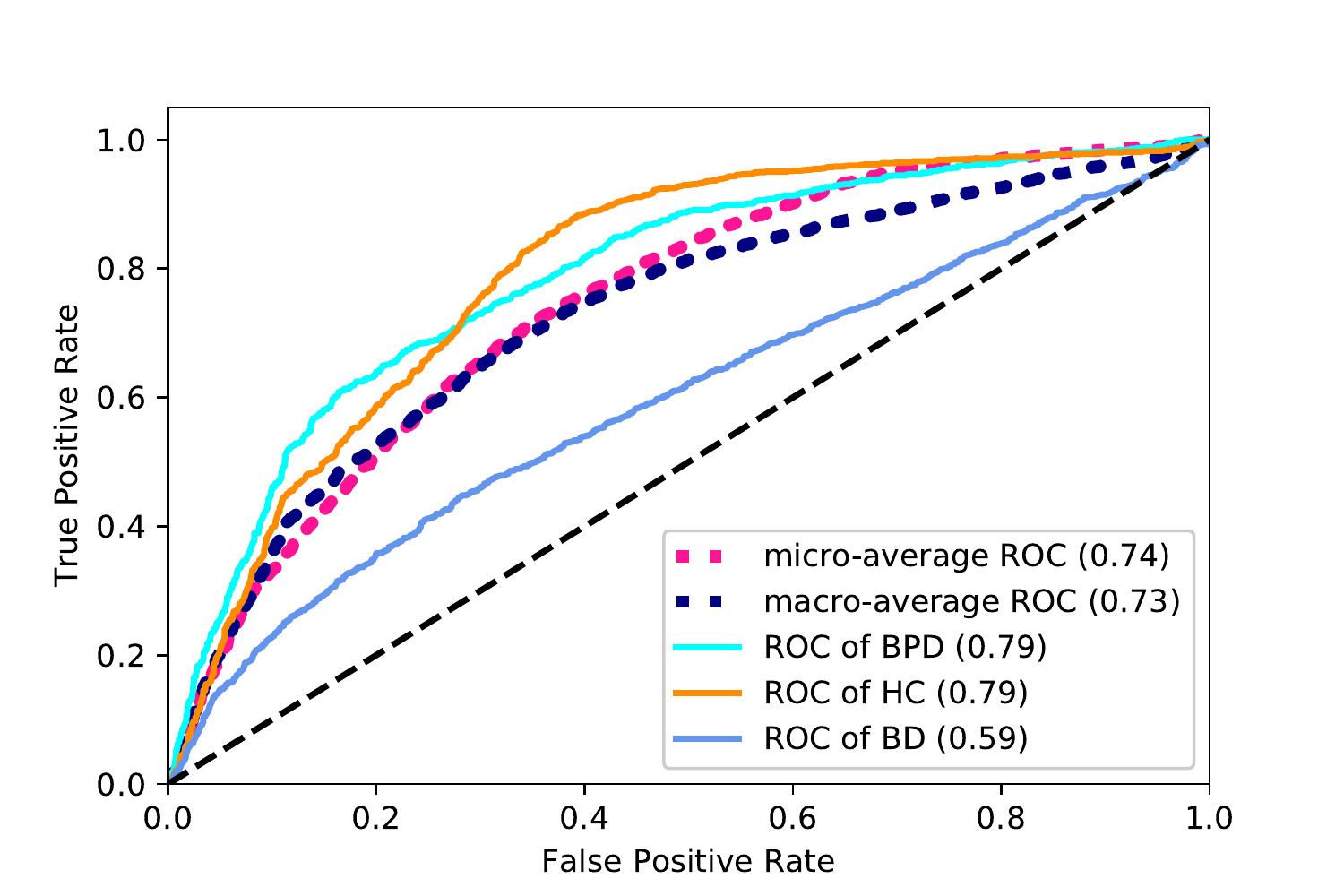}
\caption{Naive classification model}
\end{subfigure}
\caption{Receiver operating characteristic curves of the missing-response-incorporated signature-based classification model (MRSCM, level 2) and the naive classification model.}
\label{fig:roc}
\end{figure}

The receiver operating characteristic curves for three clinic groups from both methods are plotted in Figure \ref{fig:roc} with corresponding areas under curve (auc) recorded in the brackets. The MRSCM (level 2) has better ability in identifying all three diagnostic groups in terms of auc. Both methods have their lowest auc from ROC of bipolar group, which implies it is more likely for bipolar participants to be misplaced into the other two groups. 

\begin{center}
    \begin{table}[!htbp]
\caption {f1 scores for group classification using the missing-response-incorporated signature-based classification model (MRSCM, level 2) and the naive classification model.} \label{tab:f1} 
\centering
\vspace{0.3cm}
\begin{tabular}{cccc}
\toprule[1.5pt]
\midrule
Model & BD &  HC &  BPD \\  
\midrule
MRSCM (level 2)  
 & 59.2\% & 75.2\% &  60.8\%\\
Naive classification model & 50.7\% & 66.3\%   & 46.4\%\\
\midrule
\bottomrule[1.25pt]
\end{tabular}
\end{table}
\end{center}

\subsubsection{Spectrum analysis}
\label{sec:density1}
\begin{figure}[h!]
  \centering
\begin{subfigure}[b]{0.33\linewidth}
\includegraphics[trim=3cm 20cm 10cm 4cm, clip,width=1.0\textwidth]{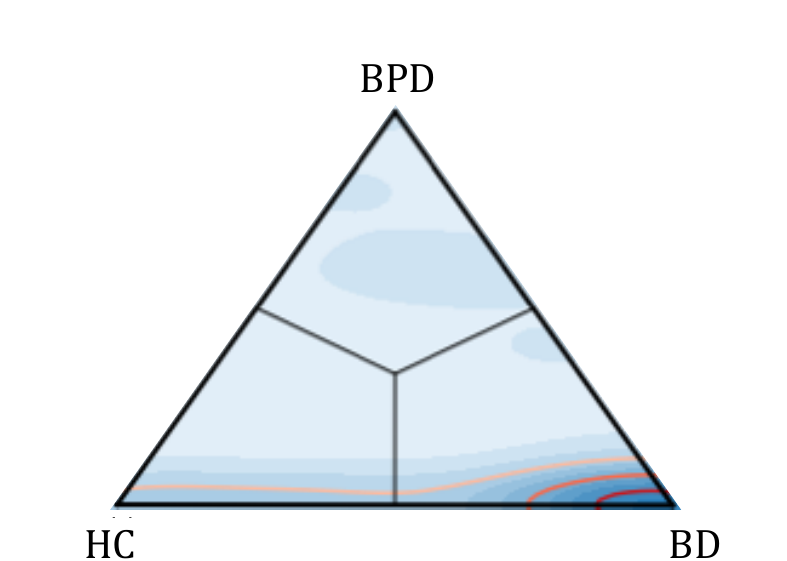}
\caption{BD participants.}
\end{subfigure}
\begin{subfigure}[b]{0.33\linewidth}
\includegraphics[trim=3cm 20cm 10cm 4cm, clip,width=1.0\textwidth]{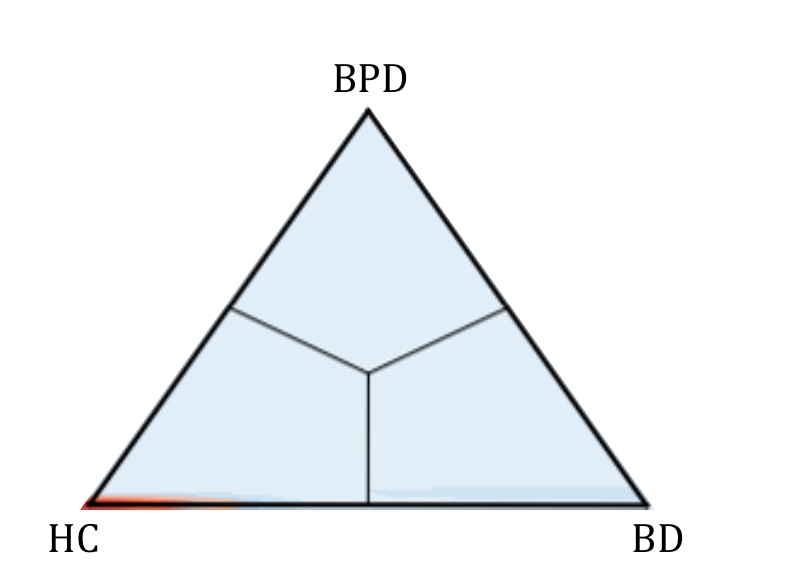}
\caption{HC participants.}
\end{subfigure}
\begin{subfigure}[b]{0.33\linewidth}
\includegraphics[trim=3cm 20cm 10cm 4cm, clip,width=1.0\textwidth]{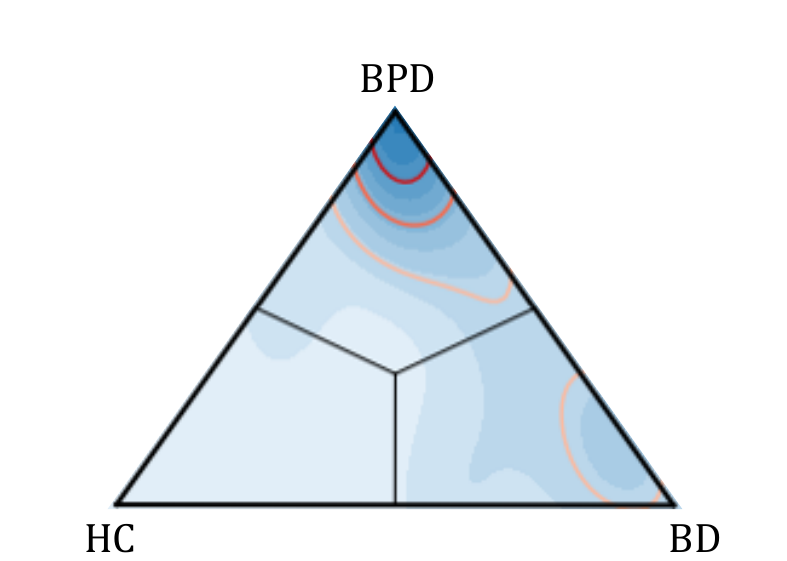}
\caption{BPD participants.}
\end{subfigure}
\\
\caption{Density plots for the predicted diagnosis from MRSCM (level 2): darker blue areas indicate higher density values, i.e., events that are more likely to happen, and vice versa; red lines indicate the 75\% (the lightest red), 50\%, 25\% (the darkest red) boundaries of density contours, i.e., the events within the area enclosed by the 75\% contour line is with probability 75\% to happen.}
 \label{fig:triangles}
\end{figure}

In Figure \ref{fig:triangles}, the triangle spectrum of the predicted diagnosis from MRSCM (level 2) are plotted. In each of the plots, the regions of highest density of participants are located in the correct corner of the triangle. The greatest consistency is with the healthy participants. Meanwhile, the probabilities of misdiagnosis to other groups can be measured by comparing the distances to the other two vertices to the distance to the right vertex. For instance, one can deduce from Plot (b) that the likelihood of misplacing healthy participants into the borderline group is very low. Plot (c) shows the other way around: BPD participants are unlikely to be misidentified as healthy control. Plot (a) shows that the bipolar patient can be misidentified as healthy control with relatively high probability and as BPD with relatively low probability.

\subsection{Prediction of the ASRM/QIDS scores of an individual participant}
Prediction of the ASRM/QIDS scores of each participant based on his/her own clinic group, are tested on bootstrapping and summarised in Table \ref{tab:accuracy_prediction}. Compared to the naive method, the model based on signature features (truncated at level 2) extracted after handling missing data is with the higher accuracy across all different clinic groups. Among the three clinic groups, Both models have their best performance in predicting the ASRM/QIDS scores for healthy controls and worst performance in predicting the ASRM/QIDS scores for borderline personality disorder. 

\begin{center}
    \begin{table}[!htbp]
\caption {\label{tab:accuracy_prediction} The average accuracy for ASRM/QIDS state prediction using the missing-response-incorporated signature-based predictive model (stateMRSPM, level 2) and the naive predictive model.}
\centering
\vspace{0.3cm}
\begin{tabular}{ccccccc}
\toprule[1.5pt]
\midrule
\multirow{2}{*}{Model}&\multicolumn{2}{l}{BD}& \multicolumn{2}{l}{HC} & \multicolumn{2}{l}{BPD}\\
&ASRM & QIDS &ASRM & QIDS &ASRM & QIDS \\
\midrule
stateMRSPM (level 2)  &  70.6\% & 64.8\%  & 79.9\% &  78.9\% &  65.2\%   &  60.2\%\\
Naive predictive model &61.7\%   & 59.5\% &70.1\% &71.6\% & 58.6\% &  55.6\%\\
\midrule
\bottomrule[1.25pt]
\end{tabular}
\end{table}
\end{center}

\begin{figure}[h!]
  \centering
\begin{subfigure}[b]{0.47\linewidth}
\includegraphics[width=\linewidth]{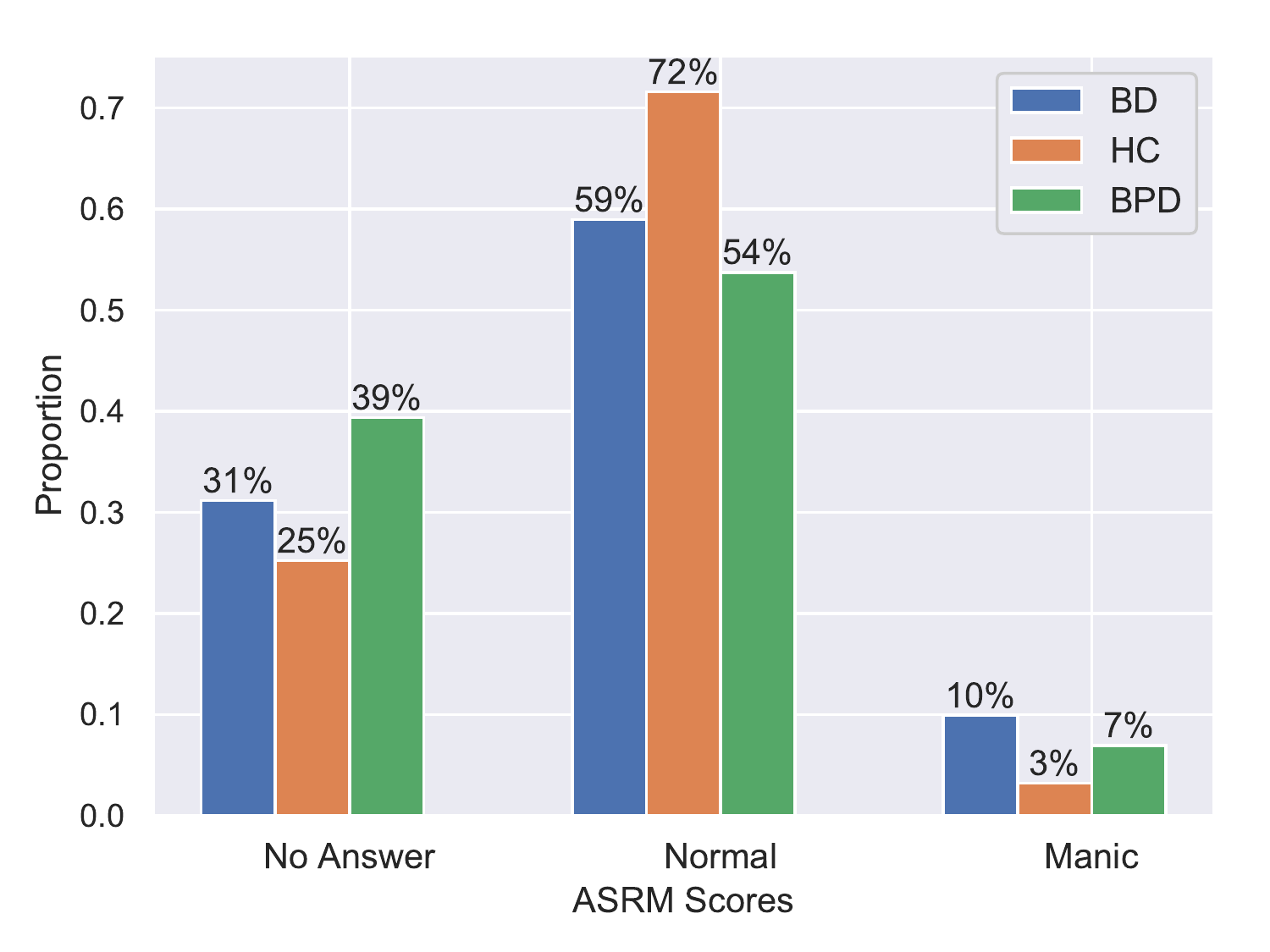}
\caption{Proportion of ASRM states}
\end{subfigure}
\begin{subfigure}[b]{0.47\linewidth}
\includegraphics[width=\linewidth]{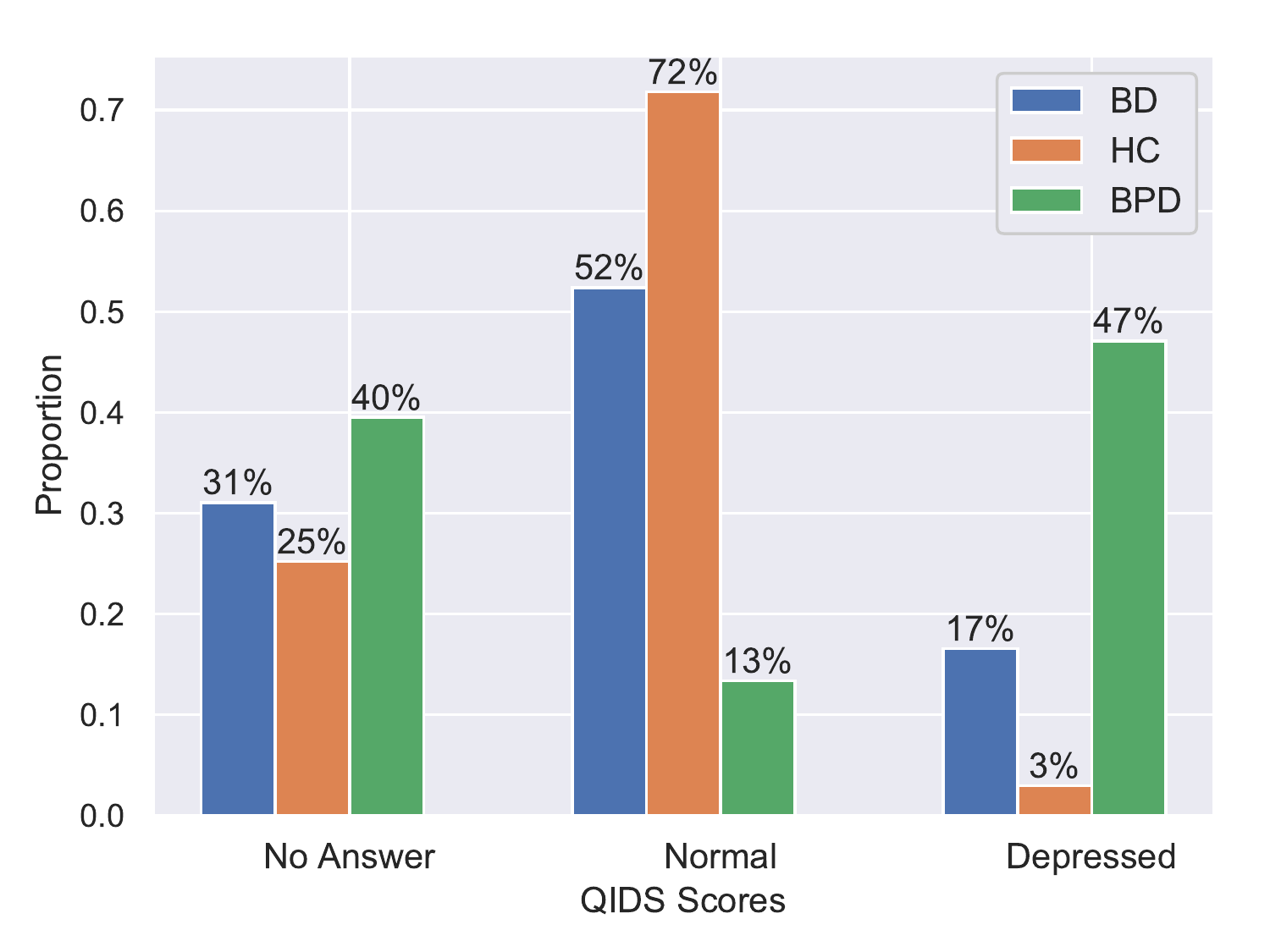}
\caption{Proportion of QIDS states}
\end{subfigure}
\caption{Bar charts: the proportion of ASRM/QIDS states of the three clinic groups, where the total number of ASRM or QIDS questionnaires for BD/HC/BPD are 620, 880, 960.}
\label{fig:trun_barchart}
\end{figure}

Figure \ref{fig:trun_barchart} demonstrates the proportions of ASRM/QIDS scores of each of the three clinic groups. HC participants rarely experienced either mania or depression in their entire study and were least likely to have missing responses. By contrast, BPD patients were most likely to have missing responses. Figure \ref{fig:triangles3_true} visualises the densities of the ASRM/QIDS states for three clinic groups using the 3-dimensional equilateral triangle surface.

\begin{figure}[h!]
  \centering
\begin{subfigure}[b]{0.33\linewidth}
\includegraphics[trim=2cm 16.7cm 8.0cm 3.5cm, clip,width=1.0\textwidth]{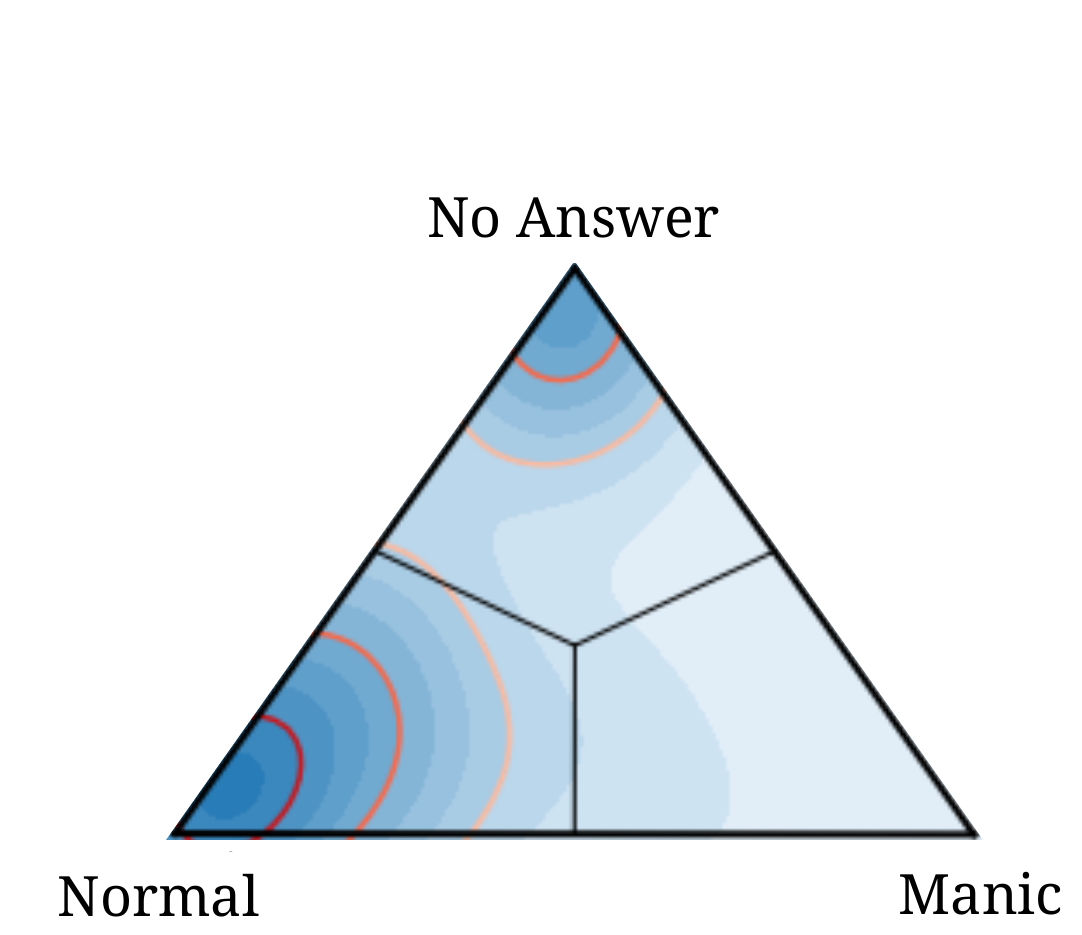}
\caption{ASRM states of BD.}
\end{subfigure}
\begin{subfigure}[b]{0.33\linewidth}
\includegraphics[trim=2cm 18cm 8.0cm 3.5cm, clip,width=1.0\textwidth]{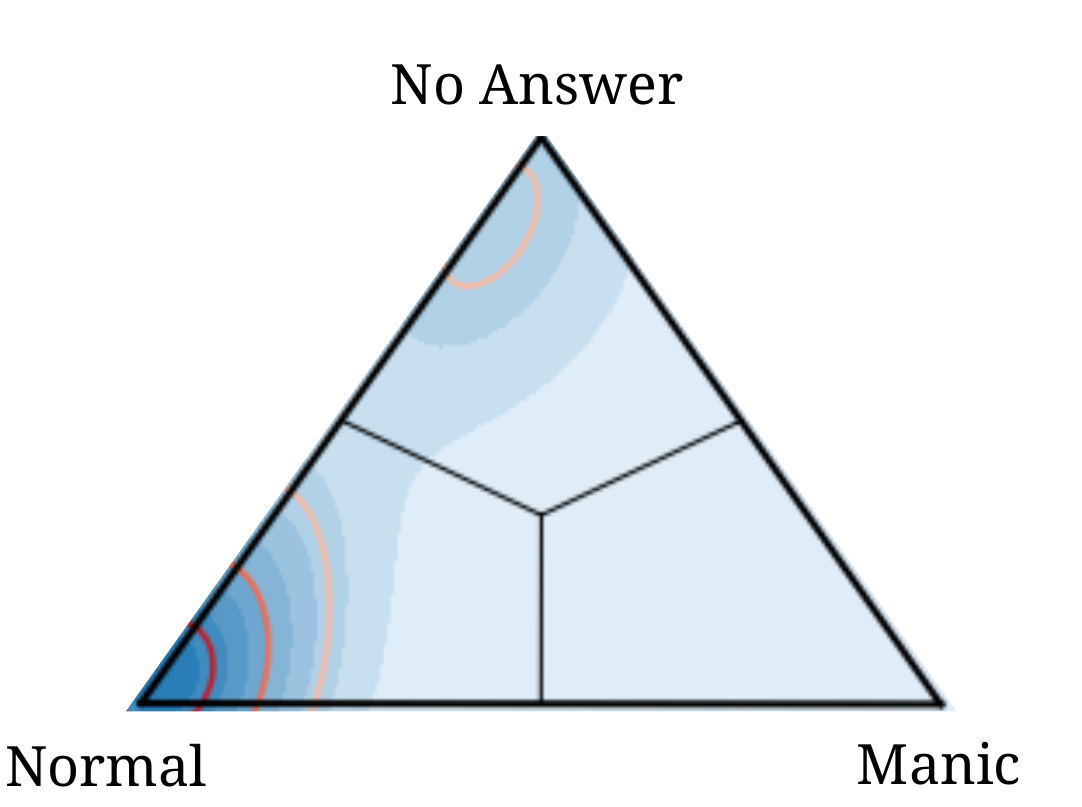}
\caption{ASRM states of HC.}
\end{subfigure}
\begin{subfigure}[b]{0.33\linewidth}
\includegraphics[trim=2cm 18cm 8.0cm 3.5cm, clip,width=1.0\textwidth]{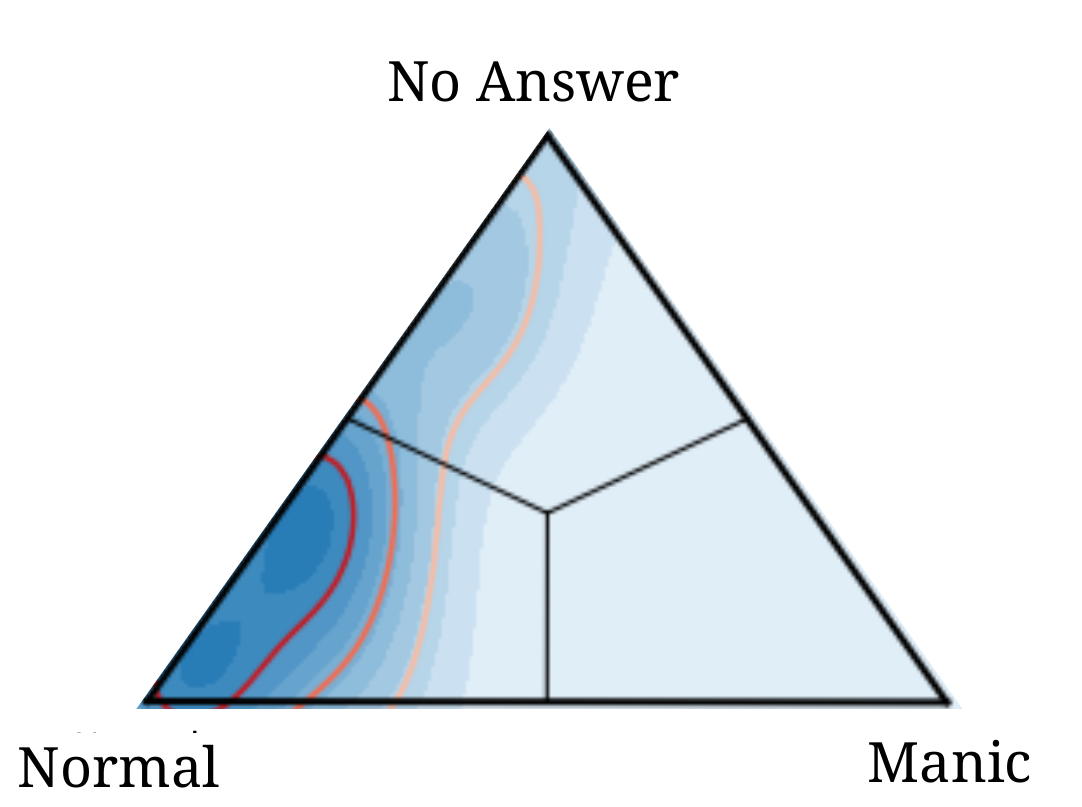}
\caption{ASRM states of BPD.}
\end{subfigure}
\\
\begin{subfigure}[b]{0.33\linewidth}
\includegraphics[trim=2cm 18cm 8.0cm 3.5cm, clip,width=1.0\textwidth]{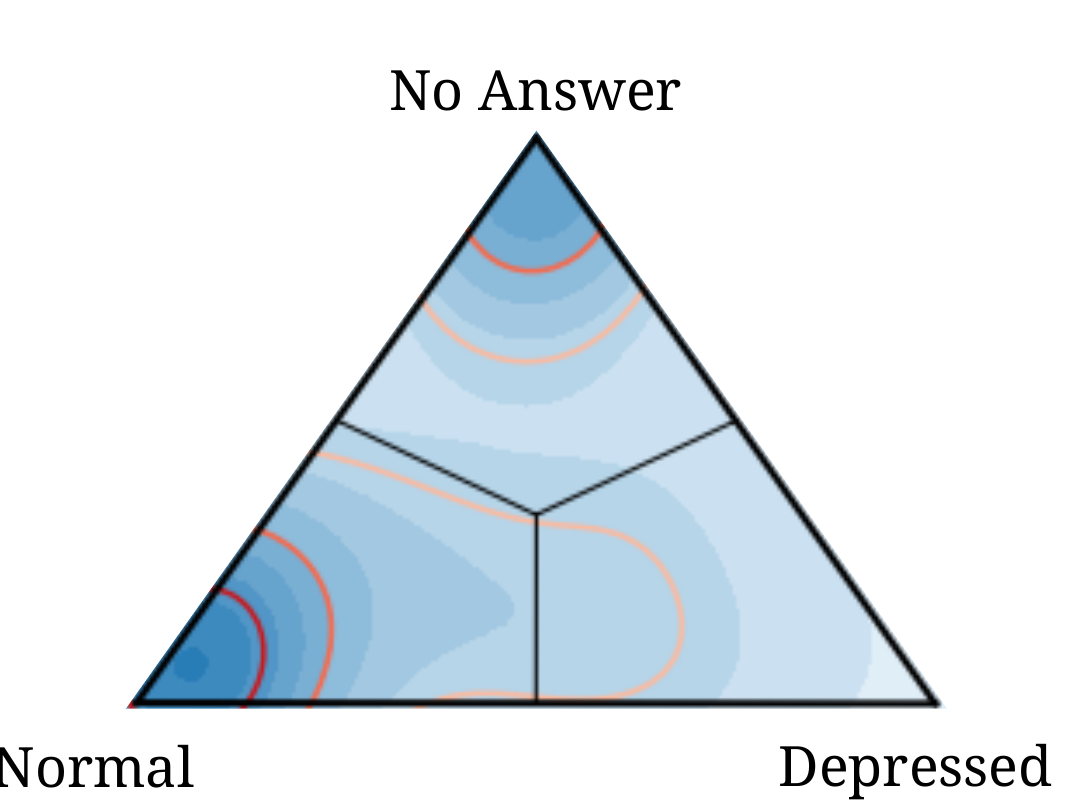}
\caption{QIDS states of BD.}
\end{subfigure}
\begin{subfigure}[b]{0.33\linewidth}
\includegraphics[trim=2cm 18cm 8.0cm 3.5cm, clip,width=1.0\textwidth]{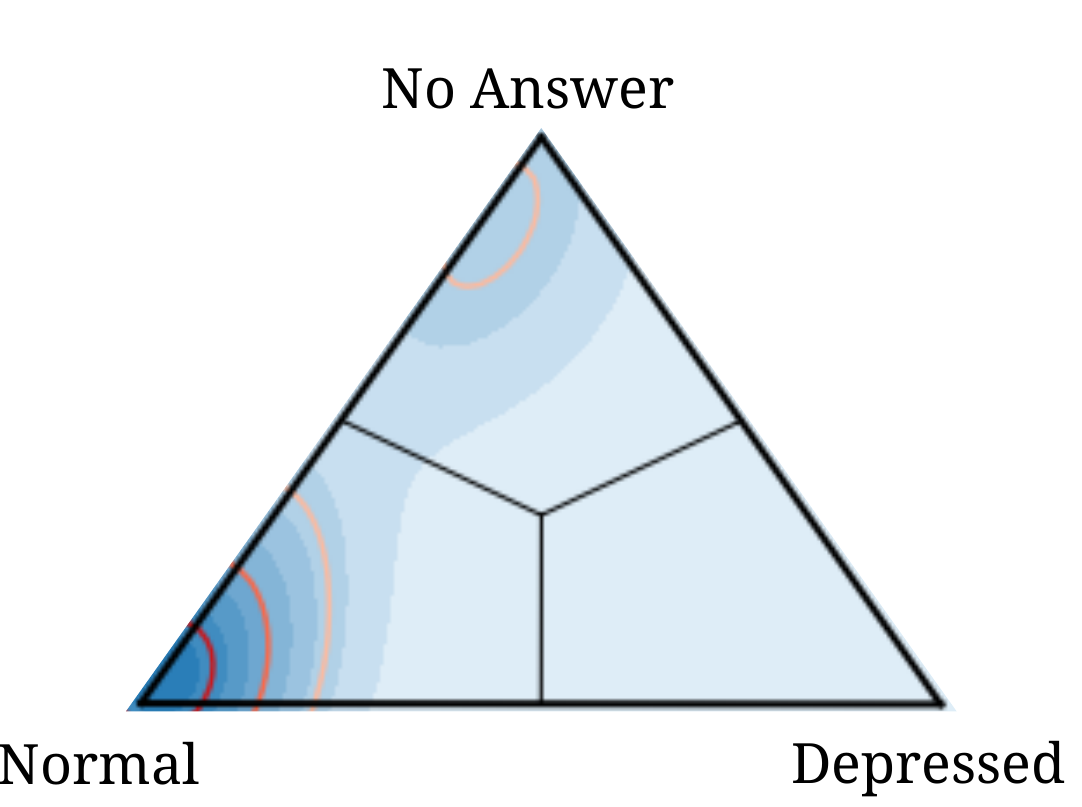}
\caption{QIDS states of HC.}
\end{subfigure}
\begin{subfigure}[b]{0.33\linewidth}
\includegraphics[trim=2cm 18cm 8.0cm 3.5cm, clip,width=1.0\textwidth]{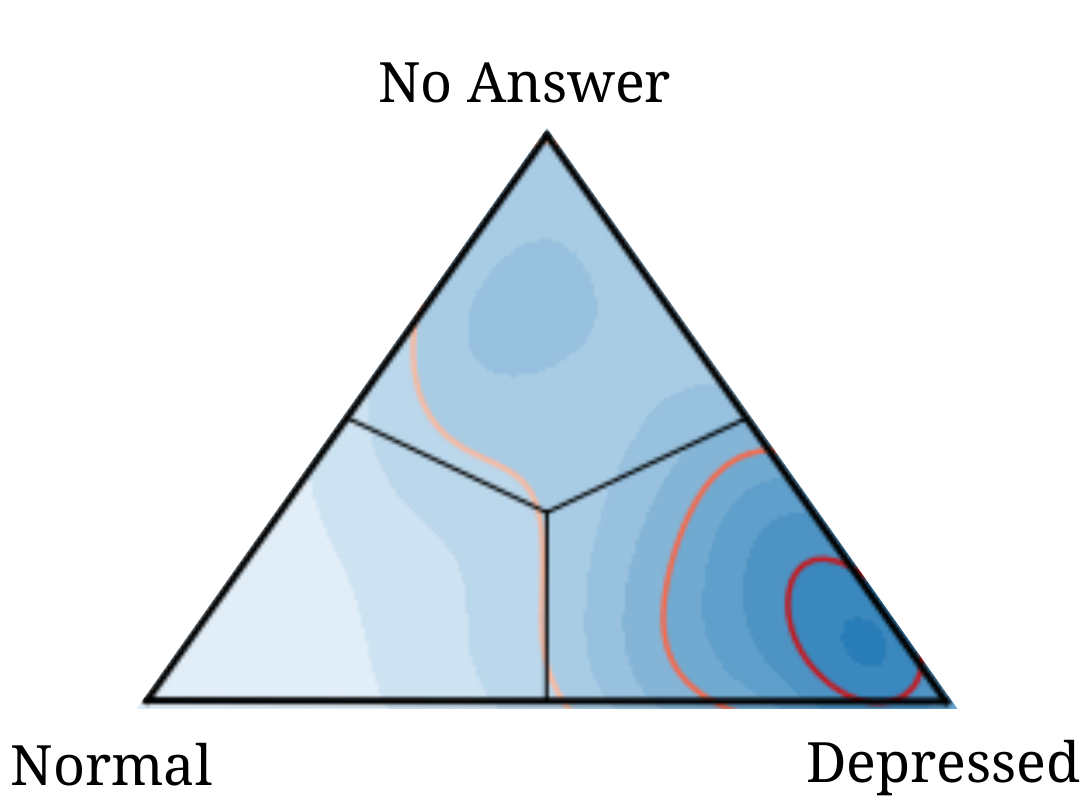}
\caption{QIDS states of BPD.}
\end{subfigure}
\caption{Density plots of true ASRM/QIDS states for three clinic groups: darker blue indicates higher density value and vice versa; red lines indicate the 75\% (the lightest red), 50\%, 25\% (the darkest red) boundaries of density contours.}
\label{fig:triangles_true}
\end{figure}

\begin{figure}[h!]
  \centering
\begin{subfigure}[b]{0.33\linewidth}
\includegraphics[trim=3cm 20cm 10cm 4cm, clip,width=1.0\textwidth]{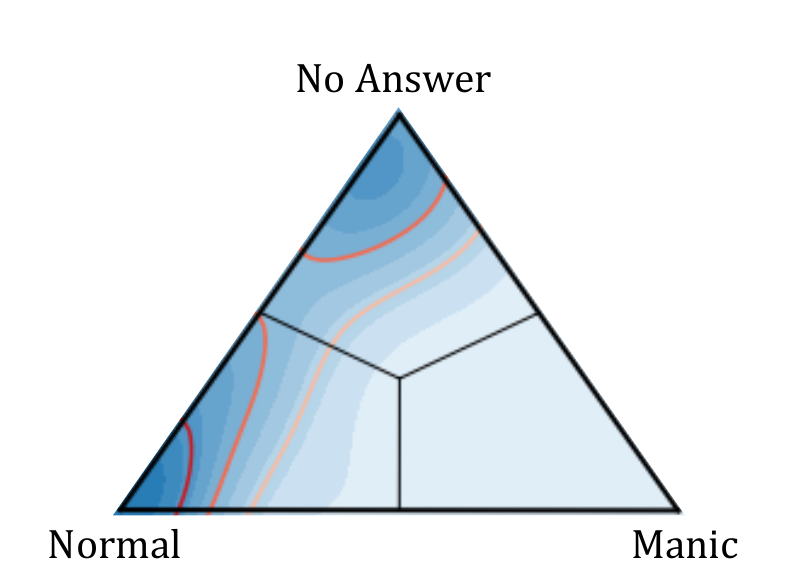}
\caption{ASRM states of BD.}
\end{subfigure}
\begin{subfigure}[b]{0.33\linewidth}
\includegraphics[trim=3cm 20cm 10cm 4cm, clip,width=1.0\textwidth]{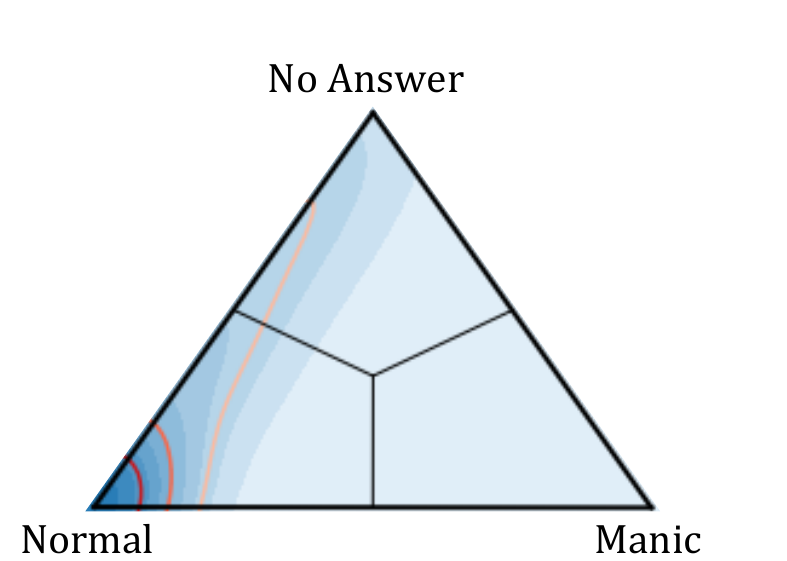}
\caption{ASRM states of HC.}
\end{subfigure}
\begin{subfigure}[b]{0.33\linewidth}
\includegraphics[trim=3cm 20cm 10cm 4cm, clip,width=1.0\textwidth]{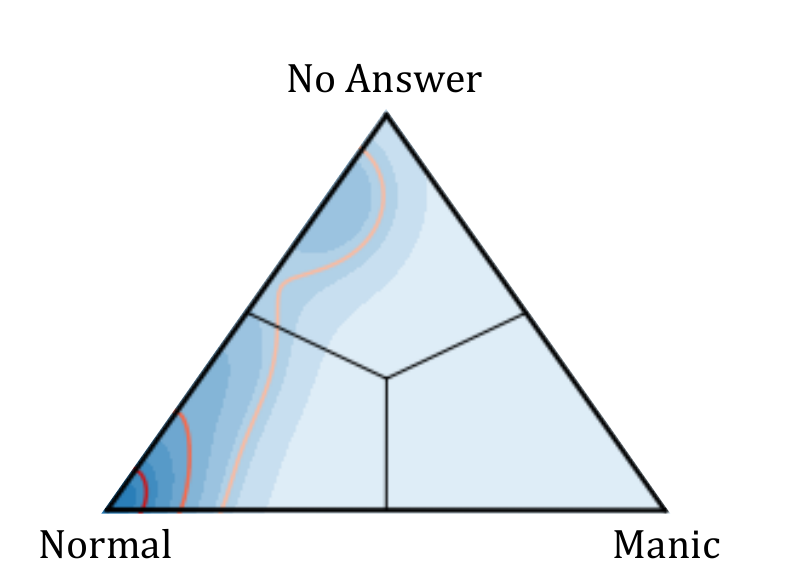}
\caption{ASRM states of BPD.}
\end{subfigure}
\\
\begin{subfigure}[b]{0.33\linewidth}
\includegraphics[trim=3cm 20cm 9.5cm 4cm, clip,width=1.0\textwidth]{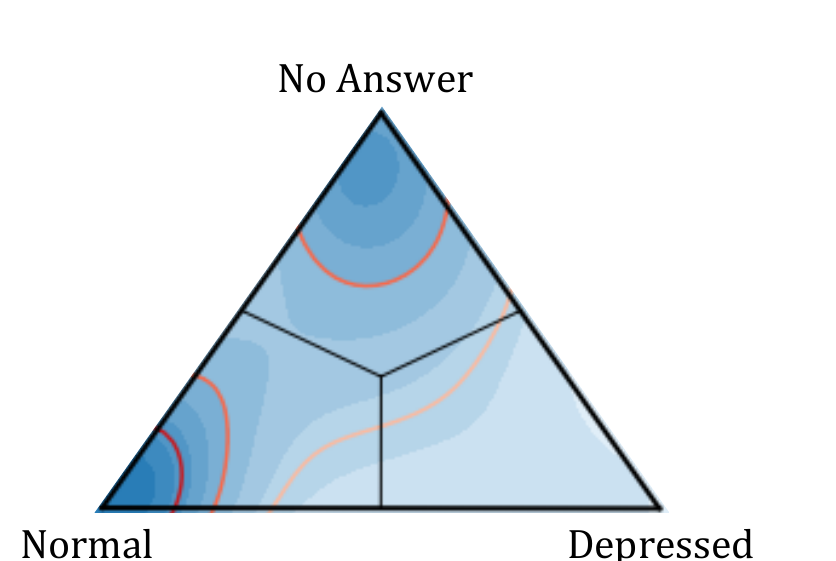}
\caption{QIDS states of BD.}
\end{subfigure}
\begin{subfigure}[b]{0.33\linewidth}
\includegraphics[trim=3cm 19.5cm 9.5cm 4cm, clip,width=1.0\textwidth]{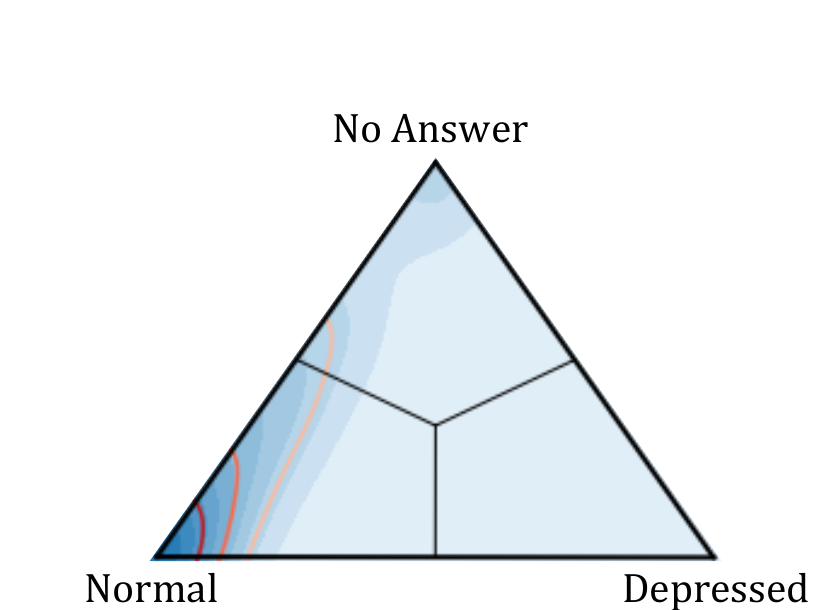}
\caption{QIDS states of HC.}
\end{subfigure}
\begin{subfigure}[b]{0.33\linewidth}
\includegraphics[trim=3cm 20cm 9.5cm 4cm, clip,width=1.0\textwidth]{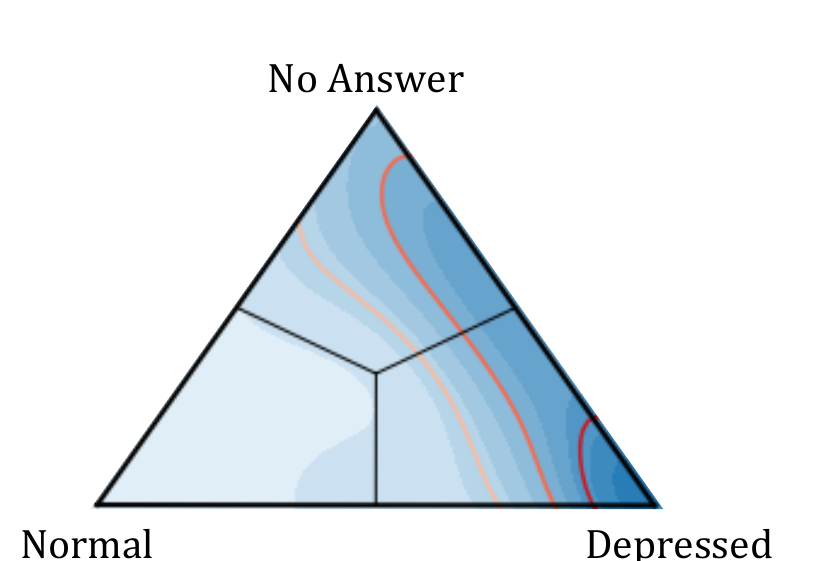}
\caption{QIDS states of BPD.}
\end{subfigure}
\caption{Density plots of the predicted ASRM/QIDS states for three clinic groups using the missing-response-incorporated signature-based predictive method (level 2): darker blue indicates higher density value and vice versa; red lines indicate the 75\% (the lightest red), 50\%, 25\% (the darkest red) boundaries of density contours}
\label{fig:triangles3}
\end{figure}

Figure \ref{fig:triangles3} demonstrates the density spectrum of the predicted ASRM/QIDS states for each clinic group using the missing-response-incorporated signature-based predictive model (level 2), which is roughly consistent with Figure \ref{fig:triangles_true}. Recall that for ASRM (resp. QIDS), only 5 future observations were predicted for each participant and therefore used for computing the proportion (i.e., density). The difference between Figure \ref{fig:triangles3} and Figure \ref{fig:triangles_true} comes from the mixed effects  of the prediction error and limited number of observations being predicted for each participant.

\subsection{Score predictions}

MAEs are recorded in Table \ref{tab:conditional2} for predicting the next reported score using scoreMRSPM (level 2) and the naive model. The worst performances were to predict ASRM scores for bipolar patients and to predict future QIDS scores for patients of borderline personality disorder. Table \ref{tab:conditional} reports the accuracy and MAE for predicting the future severity of symptoms using signature-based model. 

\begin{center}
    \begin{table}[!htbp]
\caption {\label{tab:conditional2} MAE for ASRM/QIDS score prediction using the missing-response-incorporated signature-based predictive method (scoreMRSPM, level 2) and  the naive predictive model.}
\centering
\vspace{0.3cm}
\begin{tabular}{ccccccc}
\toprule[1.5pt]
\midrule

\multirow{2}{*}{Model}&\multicolumn{2}{l}{BD}& \multicolumn{2}{l}{HC} & \multicolumn{2}{l}{BPD}\\
&ASRM & QIDS &ASRM & QIDS &ASRM & QIDS \\
\midrule
scoreMRSPM (level 2)  &  2.386& 3.437& 0.826 & 1.532 & 2.117  &  3.745\\
 Naive predictive model &3.286  &4.600&1.137 &1.899 & 2.571  &  4.671\\
 \midrule
\bottomrule[1.25pt]
\end{tabular}
\end{table}
\end{center}

\begin{center}
    \begin{table}[!htbp]
\caption {\label{tab:conditional} The average accuracy and MAE  for the prediction of severity of symptoms using the missing-response-incorporated signature-based predictive method.}
\centering
\vspace{0.3cm}
\begin{tabular}{lllllll}
\toprule[1.5pt]
\midrule

\multirow{2}{*}{Measure}&\multicolumn{2}{l}{BD}& \multicolumn{2}{l}{HC} & \multicolumn{2}{l}{BPD}\\
&ASRM & QIDS &ASRM & QIDS &ASRM & QIDS \\
\midrule
Accuracy   &  74.3\% & 76.4\%  & 95.8\% &  95.0\% & 82.4\%   &  69.8\%\\
 MAE  &  1.046& 0.685& 0.191 & 0.139 & 0.625  &  0.794\\
 \midrule
\bottomrule[1.25pt]
\end{tabular}
\end{table}
\end{center}

\section{Discussion}
\label{sec:discussion}

This paper introduces the missing-response-incorporated signature-feature-based random forest models and have them tested on the paired ASRM/QIDS data. Comprising at least 25\% of the whole dataset (Figure \ref{fig:trun_barchart}), the missing response is a remarkable and unreplaceable component and cannot be simply ignored. By integrating the informative and non-redundant resource from the missing responses,  the empowered signature features can be utilised with different machine learning models on various datasets containing missing information, either to identify the category membership based on the observed characteristics of the data or to predict the future from the past of an evolving system generated by the data. 

The missing-response-incorporated signature-based classification model is superior to the commonly used metric (the naive classification model) in differentiating the three clinic groups, i.e., bipolar disorder, healthy control and borderline personality disorder. Although signature-based methods were adopted by Perez et al \cite{arribas2018signature} to improve the existing results that used neuroimaging \cite{sato2012neuroimaging} or verbal fluency \cite{costafreda2011verbal}, our analysis provides a benchmark at the same accuracy level while utilising data of much smaller dimensions and missing responses.  

Spectrum analysis showed the overlap between the BD and HC groups in Figure \ref{fig:triangles}, which is consistent with the analysis in \cite{arribas2018signature} and with clinical experience. However, we found a much clearer differentiation between clinical groups than previous work \cite{arribas2018signature} suggesting that the inclusion of missing data added useful information. 

The prediction of future ASRM/QIDS states is notable. To our knowledge this is the first time that non-response status has been considered alongside normal status and manic/depressed status.  Borderline personality disorder tends to miss the tasks with slightly higher probability for both self-reported assessments than the other two groups, suggesting that some group-dependent feature is concealed in missing responses. 

Most of previous literature like \cite{arribas2018signature} and \cite{palmius2019geographic} were interested in predicting future scores. They did not use the missing responses as part of their information. In our analysis, the inclusion of the missing response information significantly improved our predictions. The performance (first row in Table \ref{tab:conditional2}) in general outperforms the one in \cite{palmius2019geographic} in terms of mean absolute error. Unlike models that need to impose population-level distribution and parameters \cite{busk2020forest}, our proposed model offers a unified, missing-response-incorporated and non-parametric approach that is able to capture the nature of the evolving system which generates data streams. 

For the naive models, both the performance of ASRM/QIDS-state prediction (Table \ref{tab:accuracy_prediction}) and the one of diagnostic group classification (Table \ref{tab:f1}) for BPD patients are the worst among the three clinic groups. The f1 score for classification was even below 0.5. For this task, the confusion matrix (Plot (b) in Figure \ref{fig:confusion}) also shows that about 40\% of the BPDs were misclassified as BD patients. The poor performance alerted the unreliability of this naive metric in identifying BPD patients and predicting their moods. On the other hand, by incorporating extra valuable information like missing responses into features, the signature-based model lifted the f1 score for identifying BPD patients to above 60\%, with less than one third BPD patients being misclassified as BD. This demonstrates the ability of the missing-response-incorporated signature features to capture and learn the inherent difference in mood instability between BPD and BD. Even though, the performance of ASRM/QIDS-state prediction for BPD patients from signature-based model (the last two column in Table \ref{tab:accuracy_prediction}) is still the worst among the three clinic groups. The poor performance is again consistent with the results from \cite{arribas2018signature}. This consistency may be a result of the unpredictable nature in BPD, or due to the bias from the same database that is used by both \cite{arribas2018signature} and us. 

\subsection{Limitations and implications}
The missing-response-incorporated signature-based features offer a systematic approach to the analysis of longitudinal self-reported mood data with the presence of non-randomly distributed missing values. It can be easily utilised with various machine learning methods for classification and prediction tasks on other databases containing missing information. The reasons for the moderate accuracies using MRSF are four-fold: the full potential of signature features is hindered by the small dataset, the proposed feature extraction method might not be the optimal, ASRM/QIDS data was analysed on the overall-score level instead of on the question-score level, and the diversity within the same clinic group was not considered for the prediction task. In the future, we would prioritise on two explorations: assessing our proposed method on different mental health datasets, and adjusting MRSF to the “optimal” signature-based feature by adding reasonable metrics/transformations which account for different attributes. We are also interested in addressing the inter-group difference on a much larger dataset of ASRM and QIDS responses by building independent or dependent missing-response-incorporated signature-based predictive models on clusters within the same group.  
\section*{Data, code and materials}
As the data were collected pre-GDPR and contained sensitive personal data, it cannot be placed into a publicly accessible repository.


The codes can be found through GitHub repository via \url{https://github.com/yuewu57/mental_health_AMoSS}.

\section*{Acknowledgements}
YW and TJL would acknowledge Alan Turing Institute for funding this work through EPSRCgrant EP/N510129/1 and EPSRC through the project EP/S2026347/1, titled ‘Unparameterised multi-modal data, high order signature, and the mathematics of data science’.  KEAS is supported by the NIHR Oxford Health Biomedical Research Centre. \emph{‘The views expressed are those of the authors and not necessarily those of the NHS, the NIHR or the Department of Health’.}  

\bibliographystyle{unsrt}  


\begin{thebibliography}{1}


\bibitem{altman1997asrm}
Altman, E.G., Hedeker, D., Peterson, J.L. and Davis, J.M.,
\newblock  The Altman self-rating mania scale. 
\newblock {\em Biological psychiatry}, 42.10 (1997): 948-955.

\bibitem{apa2013dsm}
American Psychiatric Association,
\newblock Diagnostic and statistical manual of mental disorders.
\newblock {\em BMC Medicine}, 17 (2013): 133-137.

\bibitem{arribas2018signature}
Arribas, I.P., Goodwin, G.M., Geddes, J.R., Lyons, T. and Saunders, K.E., 
\newblock  A signature-based machine learning model for distinguishing bipolar disorder and borderline personality disorder. 
\newblock {\em Translational psychiatry}, 8.1 (2018): 274.


\bibitem{bopp2010longitudinal}
Bopp, J.M., Miklowitz, D.J., Goodwin, G.M., Stevens, W., Rendell, J.M. and Geddes, J.R.,
\newblock The longitudinal course of bipolar disorder as revealed through weekly text messaging: a feasibility study.
\newblock {\em  Bipolar disorders}, 12.3 (2010): 327-334.


\bibitem{busk2020forest}
Busk, J., Faurholt-Jepsen, M., Frost, M., Bardram, J.E., Kessing, L.V. and Winther, O., 
\newblock Forecasting Mood in Bipolar Disorder From Smartphone Self-assessments: Hierarchical Bayesian Approach.
\newblock {\em  JMIR mHealth and uHealth},8.4 (2020): e15028.

\bibitem{costafreda2011verbal}
Costafreda, S.G., Fu, C.H., Picchioni, M., Toulopoulou, T., McDonald, C., Kravariti, E., Walshe, M., Prata, D., Murray, R.M. and McGuire, P.K., 
\newblock Pattern of neural responses to verbal fluency shows diagnostic specificity for schizophrenia and bipolar disorder. 
\newblock {\em  BMC psychiatry}, 11.1 (2011): 18.

\bibitem{faurholt2015daily}
Faurholt-Jepsen, M., Frost, M., Ritz, C., Christensen, E.M., Jacoby, A.S., Mikkelsen, R.L., Knorr, U., Bardram, J.E., Vinberg, M. and Kessing, L.V., 
\newblock Daily electronic self-monitoring in bipolar disorder using smartphones–the MONARCA I trial: a randomized, placebo-controlled, single-blind, parallel group trial. \newblock {\em Psychological medicine}, 45.13 (2015): 2691-2704.

\bibitem{faurholt2019guideline}
Faurholt-Jepsen, M., Geddes, J.R., Goodwin, G.M., Bauer, M., Duffy, A., Kessing, L.V. and Saunders, K.,
\newblock Reporting guidelines on remotely collected electronic mood data in mood disorder (eMOOD)—recommendations.
\newblock {\em Translational psychiatry}, 45.13 (2019): 1-10.

\bibitem{gaham2013sparse}
Graham, B.,
\newblock Sparse arrays of signatures for online character recognition.
\newblock {\em arXiv preprint},arXiv:1308.0371.
 


\bibitem{goodday2008monitor}
Goodday, S.M., Atkinson, L., Goodwin, G., Saunders, K., South, M., Mackay, C., Denis, M., Hinds, C., Attenburrow, M.J., Davies, J. and Welch, J.,
\newblock The true colours remote symptom monitoring system: a decade of evolution.
\newblock {\em Journal of Medical Internet Research}, 22.1 (2020): e15188.

\bibitem{hairer2013kpz}
Hairer, M.,
\newblock Solving the KPZ equation.
\newblock {\em Annals of Mathematics}, (2013): 559-664.

\bibitem{hairer2014regularity}
Hairer, M.,
\newblock A theory of regularity structures.
\newblock {\em Inventiones mathematicae},  198.2 (2014): 269-504.

\bibitem{levin2013past}
Levin, D., Lyons, T. and Ni, H.,
\newblock Learning from the past, predicting the statistics for the future, learning an evolving system.
\newblock {\em arXiv preprint},arXiv:1309.0260.

\bibitem{little2019missing}
Little, R.J. and Rubin, D.B.,
\newblock {\em Statistical analysis with missing data}, John Wiley \& Sons, 793 (2019).

\bibitem{lyons2002rough}
Lyons, T. and Qian, Z.,
\newblock {\em System control and rough paths},  Oxford University Press (2002).

\bibitem{lyons2014rough}
Lyons, T.,
\newblock Rough paths, signatures and the modelling of functions on streams.
\newblock {\em arXiv preprint},  arXiv:1405.4537.

2019. Using path signatures to predict a diagnosis of Alzheimer’s disease. PloS one, 14(9).

\bibitem{moore2019using}
Moore, P.J., Lyons, T.J., Gallacher, J. and Alzheimer’s Disease Neuroimaging Initiative, 
\newblock Using path signatures to predict a diagnosis of Alzheimer’s disease.
\newblock {\em PloS one},  14.9 (2019).


\bibitem{palmius2019geographic}
Palmius, N., Tsanas, A., Saunders, K.E.A., Bilderbeck, A.C., Geddes, J.R., Goodwin, G.M. and De Vos, M., 
\newblock Detecting bipolar depression from geographic location data. 
\newblock {\em IEEE Transactions on Biomedical Engineering},  64.8 (2016):1761-1771.

\bibitem{reizenstein2018iisignature}
Reizenstein, J. and Graham, B.,
\newblock The iisignature library: efficient calculation of iterated-integral signatures and log signatures.
\newblock {\em arXiv preprint},  arXiv:1802.08252.

\bibitem{rush2003qids}
Rush, A.J., Trivedi, M.H., Ibrahim, H.M., Carmody, T.J., Arnow, B., Klein, D.N., Markowitz, J.C., Ninan, P.T., Kornstein, S., Manber, R. and Thase, M.E.,
\newblock The 16-Item Quick Inventory of Depressive Symptomatology (QIDS), clinician rating (QIDS-C), and self-report (QIDS-SR): a psychometric evaluation in patients with chronic major depression.
\newblock {\em  JBiological psychiatry}, 54.5 (2003): 573-583.


\bibitem{sato2012neuroimaging}
Sato, J.R., de Araujo Filho, G.M., de Araujo, T.B., Bressan, R.A., de Oliveira, P.P. and Jackowski, A.P.,
\newblock Can neuroimaging be used as a support to diagnosis of borderline personality disorder? An approach based on computational neuroanatomy and machine learning.
\newblock {\em Journal of psychiatric research}, 46.9 (2012): 1126-1132.

\bibitem{tsanas2016truecolor}
Tsanas, A., Saunders, K.E.A., Bilderbeck, A.C., Palmius, N., Osipov, M., Clifford, G.D., Goodwin, G.M. and De Vos, M.,
\newblock Daily longitudinal self-monitoring of mood variability in bipolar disorder and borderline personality disorder.
\newblock {\em Journal of affective disorders}, 205 (2016): 225-233.


\bibitem{wang2019speech}
Wang, B., Liakata, M., Ni, H., Lyons, T., Nevado-Holgado, A.J. and Saunders, K.,
\newblock A Path Signature Approach for Speech Emotion Recognition.
\newblock {\em  Interspeech 2019}, ISCA (2019): 1661-1665.



\bibitem{xie2017learning}
Xie, Z., Sun, Z.,  Jin, L.,  Ni, H.,  and  Lyons, T.,
\newblock  Learning spatial-semantic context with fully convolutional recurrent network for online
handwritten Chinese text recognition.
\newblock {\em  IEEE transactions on pattern analysis and machine intelligence},   40.8 (2017): 1903-1917.

\bibitem{yang2017skeleton}
Yang, W., Lyons, T., Ni, H., Schmid, C.,  Jin, L., and  Chang, J.,
\newblock Leveraging the Path Signature for Skeleton-based Human Action Recognition.
\newblock {\em arXiv preprint},  arXiv:1707.03993.



\end{thebibliography}

\section*{Appendix}
\setcounter{table}{0}
\renewcommand{\thetable}{A\arabic{table}}

\begin{center}
    \begin{table}[!htbp]
\caption {A summary of tasks, where conPredition is short for conditional prediction.} \label{tab:tasksummary} 
\centering
\vspace{0.3cm}
\begin{tabular}{lll}
\toprule[1.5pt]
\midrule
 Task & Description & Classes or Values\\
\midrule
Classification &To identify the participant's diagnostic group   &BD/HC/ BPD \\
\midrule
State prediction  &To predict the participant's future mood state  &ASRM: No response/Normal/Manic  \\
&    &QIDS: No response/Normal/Depressed  \\
\midrule
Score prediction  & To predict the participant's future raw score &ASRM: 0-20 \\
 &given that it is not missing& QIDS: 0-27\\
 \midrule
Severity prediction  & The future severity of symptoms obtained by &ASRM: 0-4 \\
 &cut-off scores of score prediction & QIDS: 0-4\\
\midrule
\bottomrule[1.25pt]
\end{tabular}
\end{table}
\end{center}

\begin{center}
    \begin{table}[!htbp]
\caption {A summary of models, where MR is short for missing responses, RF short for random forest and conPredition short for conditional prediction.} \label{tab:modelsummary} 
\centering
\vspace{0.3cm}
\begin{tabular}{cccccc}
\toprule[1.5pt]
\midrule
 \multirow{2}{*}{ Task} & \multirow{2}{*}{ Model} & \multirow{2}{*}{ Data length} & \multicolumn{2}{c}{Feature extraction}&\multirow{2}{*}{Base model}\\
&& & MR integration & Signatures & \\
\midrule
Classification &MRSCM (level 2) & 20& Yes  & Yes &RF classifier  \\
& Naive classification model & 20 & No  & No &RF classifier \\
\midrule
State prediction &  stateMRSPM (level 2) &10 &Yes  & Yes &RF classifier   \\
&  Naive predictive model & 10&   No&No &RF classifier   \\
\midrule
Score prediction & scoreMRSPM (level 2) & 10 & Yes  & Yes&RF regressor \\
&  Naive predictive model & 10&   No&No &RF regressor   \\
\midrule
\bottomrule[1.25pt]
\end{tabular}
\end{table}
\end{center}

\end{document}